\documentclass[twocolumn,twoside,showkeys,preprintnumbers,amsmath,amssymb,secnumarabic, nofootinbib, nobibnotes,epl]{revtex4}

\usepackage{epsfig}

\usepackage{graphicx}   

\usepackage{fancyhdr}

\usepackage{pslatex}

\newcommand{\be}{\begin{equation}}
\newcommand{\ee}{\end{equation}}
\newcommand{\bear}{\begin{eqnarray}}
\newcommand{\eear}{\end{eqnarray}}
\newcommand{\sect}[1]{\setcounter{equation}{0}\section{#1}}

\begin{document}

\title{Anomalous enhancements of low-energy fusion rates in plasmas:
the role of ion momentum distributions and inhomogeneous screening}

\author{M. Coraddu$^{1,\ast}$,  Marcello~Lissia$^{1,2,\S}$,P. Quarati$^{1,3,4\ddag}$}
\affiliation{$^1$Istituto Nazionale di Fisica Nucleare
(INFN), Sezione di Cagliari, I-09042, Monserrato, Italy\\
$^2$Dipart. di Fisica dell'Universit\`a di Cagliari,
Cittadella universitaria, I-09042 Monserrato, Italy\\
$^3$Dipartimento di Fisica - Politecnico di Torino,
I-10129, Italy\\
$^4$Istituto Nazionale per la
Fisica della Materia (CNR--INFM), Sezione del Politecnico di Torino,
I-10129, Italy}

\begin{abstract}
Non-resonant fusion cross-sections significantly higher
than corresponding theoretical predictions are observed in
low-energy experiments with deuterated matrix target. Models based
on thermal effects, electron screening, or quantum-effect dispersion
relations have been proposed to explain these anomalous results:
none of them appears to satisfactory reproduce the experiments.
Velocity distributions are fundamental for the reaction rates and
deviations from the Maxwellian limit could play a central role in
explaining the enhancement. We examine two effects: an increase of
the tail of the target Deuteron momentum distribution due to the
Galitskii-Yakimets quantum uncertainty effect, which broadens the
energy-momentum relation; and spatial fluctuations of the
Debye-H\"{u}ckel radius leading to an effective increase of electron
screening. Either effect leads to larger reaction rates especially
large at energies below a few keV, reducing the discrepancy between
observations and theoretical expectations. 
\end{abstract}


\maketitle

\thispagestyle{fancy}

\setcounter{page}{1}

\sect{\label{sec:intro}   Introduction}

In the last ten years a number of different experiments with target
Deuterons absorbed in a metallic matrix have found strong
enhancements of fusion reaction rates below a few keV.The fusion of
the implanted Deuterons with the incoming ions (D$^+$ or Li$^+$) has
been observed; the  $d(d,t)p$\/  reaction has been investigated in
refs.
\cite{Yuki:98,Kasagi:02,Czerski01,Raiola:02,Raiola:02a,Bonomo:03,%
Raiola:04,Raiola:05,Raiola:06,Huke:07,Huke:08a} and the
$^{6,7}Li(d,\alpha)^{4,5}He$\/ reactions have been studied in refs.
\cite{Kasagi:04a,Kasagi:04,{Cruz:05},{Cruz:08}} finding similar
strong enhancements. Experiments with  gas targets show much weaker
enhancements, which can be explained by standard electron screening
with a potential $U_e$\/ of the same order of the adiabatic limit
$U_{ad}=28$\/ eV. In other words, the penetration through a screened
Coulomb barrier at energy $E$\/ is equivalent to that of bare nuclei
at energy $E+U_e$. The adiabatic limit is reached when the
correspondent enhancement of nuclear cross section can be explained
by the gain of the electron binding energies $U_{ad}$\/ between the
initial distant atoms and the final fused nuclei setting $U_e =
U_{ad}$\/ \cite{Salpeter:69}.

The same type of screening mechanism could reproduce
results for deuterated metal target experiments only using
an unreasonable large potential $U_e$\/ of hundreds of eV,
ten times greater than the adiabatic limit $U_{ad}$.

A tentative explanation
\cite{Bonomo:03,Raiola:04,Raiola:05,Raiola:06},
based on a simplified model of
the classical quasi-free electrons, needs
an electron screening distance of the order of the Debye length.
This approach
reproduces both the correct size of the screening potential $U_e$\/
and its dependence on the temperature: $U_e \propto
T^{1/2}$~\cite{Bonomo:03, Raiola:05}. However, this model
lacks a clear physical interpretation such as the
one for the Debye screening, which is a cooperative
effect, since
the mean-number of quasi-free
particles in the Debye sphere \cite{Ichimaru:1992}
is much smaller than one.

The thermal motion of the target atoms is another mechanism capable
of increasing the reaction rate; however, Maxwellian momentum
distribution at the experimental temperatures gives negligible
effects~\cite{Fi:03, Starostin:2003next}. The relationship between
energy and momentum of quasi particles can be  broaden by  many-body
collisions \cite{Ga:67}, then even a Maxwell-Boltzmann energy
distribution leads to a momentum distribution with an enhanced
power-law tail. Fusion processes select high-momentum particles that
are able to penetrate the Coulomb barrier and are, therefore,
extremely sensitive probes of the distribution
tail~\cite{Coraddu:1998yb,St:00,St:02,Starostin:2005PhysUsp,Lissia:2005em}.
This Quantum  Uncertainty Effect has been proposed in
\cite{Starostin:2003next,Starostin:2003next2,Starostin:2005PhysUsp,Coraddu:06,Zubarev:06,Zubarev:07}
as a possible contribution to the reaction rate enhancements.

These two mechanisms clearly cannot take into account all the
complex many-body physics in plasmas. In different contexts other
approaches exist both to fluctuations and to strong-coupled
screening ~\cite{Salpeter:69}.

 Among the over 50 different
targets (metals and insulators) where deuteron fusion reactions have
been studied, this present work focuses on the paradigmatic case of
Ta matrix to show how either of the two effects we consider reduces
the discrepancy between theoretical predictions and experimental
data. This case has been extensively studied by different
experimental groups and there exist many published data
\cite{Czerski01,Raiola:02,Raiola:02a,Bonomo:03,Huke:07,Huke:08a} in
reasonable agreement. Both effects yield qualitatively similar
results when applied to other targets, but detailed quantitative
comparisons need further work.

We review  the experimental procedure in  section \ref{sec:ExProc}
and evaluate the consequence of the Quantum  Uncertainty Effect  in
section \ref{sec:QuantumEff}. The plasma screening effects on the
reaction rate are evaluated in section \ref{sec:DefDistr}, using for
the first time the  modified Debye-H\"{u}ckel potential proposed by
Quarati and Scarfone in ref. \cite{Quarati:07}, which is an approach
introduced to study deviations from the weakly coupled plasma limit,
where the standard Debye-H\"{u}ckel screening applies. We draw our
conclusions in section \ref{sec:Conc}.

\sect{\label{sec:ExProc} Experimental procedure} 

All experiments performed to investigate the $D+D\rightarrow T+p$\/
fusion reaction, in all different target matrices, employ a
Deuterons ion beam of energy 1 keV $\leq E_b \leq 100$ keV, that is
totally absorbed by the thin foil ($\sim 0.1$\/ mm) of deuterated
target, at room temperature of about $10^\circ$C. Detectors count
the total number $N(E_b, \theta)$\/ of fusion reaction protons
emitted  in $\theta$\/ directions, and the results can be expressed
in terms of the reaction Yield of an infinitely thick target:
$Y_{exp}^{\infty} = N(E_b , \theta )/N_p$, where $N_p$\/ is the
total number of incident projectiles.

If we consider the target nuclei
at rest, the thick target reaction Yield can be
expressed as:
\be
   Y_{th}^{\infty}(E_b , \theta)\, =\, \epsilon\, n_D\, \int_0^{E_b}
                                      \sigma(E)\, \left(\frac{dE}{dx}\right)^{-1}\, dE\; ,
   \label{eq:YieldThickTargetTh}
\ee
where $n_D$\/ is the density of the target deuterons,
$\epsilon$\/ is the proton detection efficiency,
$ \sigma$(E)\/ is the fusion cross section and
$\left(\frac{dE}{dx}\right)$\/ is the energy loss for unit length,
or Stopping Power.
The effect of the thermal motion of target deuterons modifies
$Y_{th}^{\infty}$\/ in:
\be
   Y_{th}^{\infty}(E_b , \theta)\, =\, \epsilon\, n_D\, \int_0^{E_b}
                                      \frac{\langle\sigma v_{rel} \rangle}{v}\,
                                      \left(\frac{dE}{dx}\right)^{-1}\, dE\; ,
   \label{eq:TermalYieldThickTargetTh}
\ee
where
$ v_{rel} = |\mathbf v - \mathbf v_t|$\/ is the relative velocity,
$v_t$\/ and $v=\sqrt{2E/m_D}$\/ are the target and the incident particle velocities,
$E$\/ is the incident deuteron energy inside the target,
$m_D$\/ is the deuteron mass, and
\( \langle\sigma v_{rel} \rangle = \int d^3\mathbf{p}_t \Phi(\mathbf{p}_t) \sigma\, v_{rel}\; \)\/
is the thermal mean with
$ \Phi(\mathbf{p}_t)$ the distribution
of particle momentum target $\mathbf{p}_t$.
For the Stopping Power $\left(\frac{dE}{dx}\right)$\/
all the authors adopt the values reported by
Andersen and Ziegler \cite{And:08}.

In the zero-temperature approximation the
reaction cross section $\sigma_{exp}(E)$\/ can be extracted from
$Y_{exp}^{\infty}$\/ through a finite-interval numeric differentiation:
\(    \sigma_{exp}\, =\, \frac{\left(\frac{dE}{dx}\right)}{\epsilon\, n_D}\;
    \frac{Y_{exp}^{\infty}(E_b , \vartheta)\, -\,
          Y_{exp}^{\infty}(E_b - \Delta E_b , \vartheta)}{\Delta E_b} \;  \),
where $\Delta E_b$\/ is a small beam energy step between two
subsequent measurements of $Y_{exp}^{\infty}$.
If thermal effects are included, this procedure yields
$\left(\langle\sigma v_{rel} \rangle /v \right)_{exp}$\/
instead of $\sigma_{exp}$\/.
A convenient theoretical expression for the bare-nucleus cross section is:
\be
  \sigma_{th}(E_{cm})\, =\, \frac{S(E_{cm})}{E_{cm}}\, \exp \left[ - \pi \sqrt{\frac{E_G}{E_{cm}}} \right]\; ,
  \label{eq:CrosSecbareNuc}
\ee
where the exponential factor is the penetration function across the
bare repulsive Coulomb potential,
$E_G=2\mu Z_{1}^{2} Z_{2}^{2} e^{4} /\hbar^2$\/ is the Gamow energy,
$\mu$\/ is the reduced mass
$Z_{1}\, ,\, Z_{2}$\/ are the atomic numbers of the interacting nuclei and
$S(E_{cm})$\/ is the astrophysical factor. If the small thermal effects are neglected,
the center of mass energy is $E_{cm} = E_b/2$.
Static electron screening can be taken into account by introducing the electron screening potential
 $U_e$\/:
\be
  \sigma_{th}(E_{cm})\, =\, \frac{S(E_{cm})}{E_{cm}+U_e}\, \exp \left[ - \pi \sqrt{\frac{E_G}{E_{cm}+U_e}} \right]\; .
  \label{eq:CrosElScreenedNuc}
\ee

\begin{figure}
 \begin{center}

        \includegraphics[scale=0.9]{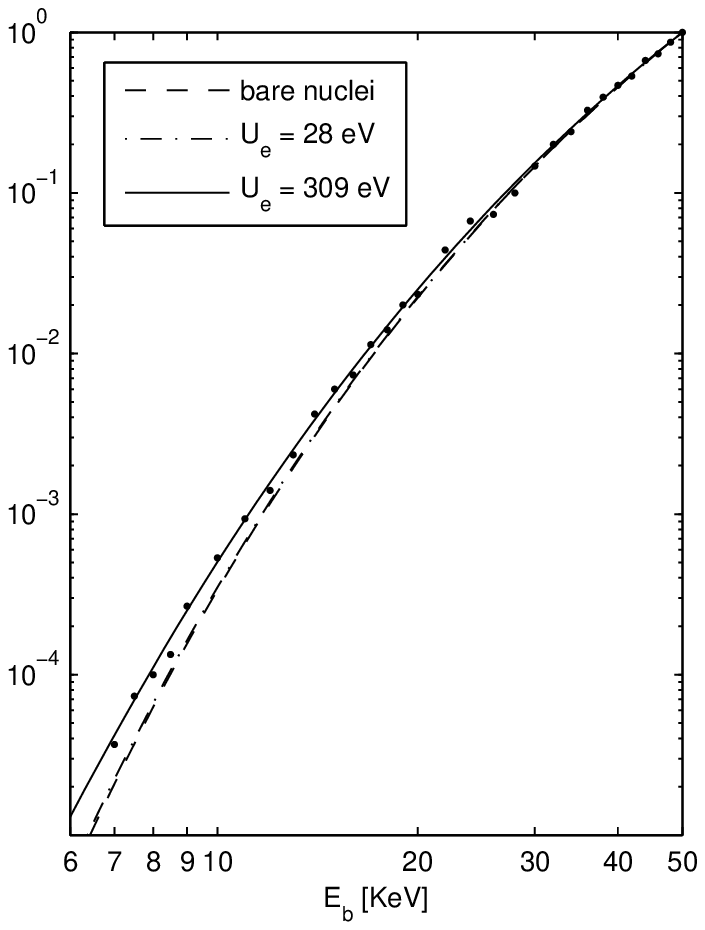}
        \includegraphics[scale=0.9]{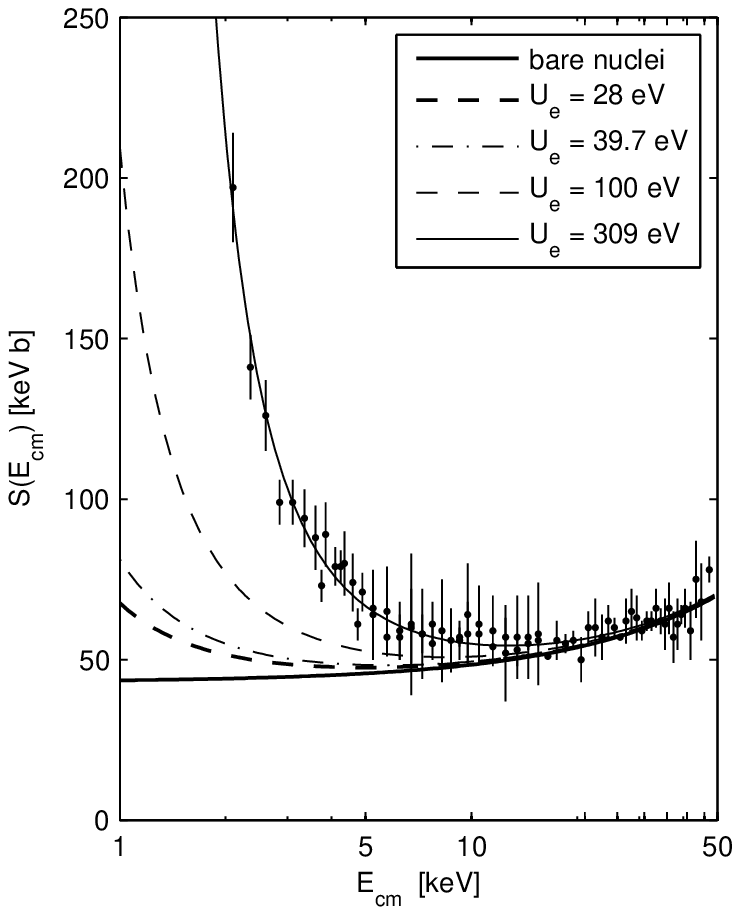}

 \end{center}
 \caption{\footnotesize Experimental results from ref. \cite{Raiola:02}.
          Upper panel: $Y_{exp}^{\infty}$\/ data points from one of the 13 runs
          weighted to extract the cross section  $\sigma_{exp}$; data normalized to
          a beam energy $E_b$\/ (Lab frame) of 50 keV; $Y_{th}^{\infty}$\/ curves
          from eq. (\ref{eq:YieldThickTargetTh}) are computed with
          eq. (\ref{eq:CrosSecbareNuc}) (bare) and  eq. (\ref{eq:CrosElScreenedNuc})
          (screened) cross sections.
          Lower panel: Astrophysical Factor data points $S_{exp}$\/ extracted from
          $Y_{exp}^{\infty}$;  bare Astrophysical Factor is assumed to be
          $S(E_{cm})=43 +\, 0.54\, E_{cm}$\/ keV b,
          by normalization with higher energy results ($E_{cm}\geq 40$ keV),
          while screened $S_{th}$\/ curves are obtained from eq. (\ref{eq:CrosElScreenedNuc}).
          Thermal effect are neglected. }
 \label{fig:Raiola2002Experiment}
 \end{figure}

Theoretical calculations  and experimental results are often
conveniently reported in terms  of the Astrophysical Factor $S(E)$:
\( S_{th,exp\,}(E_{cm})\, =\, E_{cm}\, \exp \left[ \pi
\sqrt{\frac{E_G}{E_{cm}+U_e}} \right]\, \sigma_{th,exp} \;  \),
since it depends much less strongly on energy.
 Of course, while
$S_{exp\,}(E_{cm})$\/ is extracted from $Y_{exp}^{\infty}$, the
quantity $S_{th\,}(E_{cm})$\/ depends on the value of $U_e$\/ and on
the theoretical cross section. $S_{th\,}(E_{cm})$\/ would in
principle include all many-body effects, which modify the bare cross
section in vacuum. In this work we consider, separately, only two
effects as dominant for the reaction rate: the quantum broadening
effect on the momentum distribution and the electron screening,
which we evaluate beyond the Debye-H\"{u}ckel approximation given
the plasma density inside the metal matrix.
 The case of Deuterated Tantalum target has been investigated
extensively
\cite{Czerski01,Raiola:02,Raiola:02a,Bonomo:03,Huke:07,Huke:08a}.
Measurements are reported with great detail in ref.
\cite{Raiola:02}, then we adopt the relative experimental
environment for our comparison: matrix target chemical composition
is pure Ta ($Z=73$, $A$=180.948 a.m.u.), target temperature
$T=10^\circ$C,  Ta density at room temperature $\rho$=16.65 g
cm$^{-3}$\/ ($n_{Ta} = 5.54 \cdot 10^{22}\,$ cm$^{-3}$),
stoichiometric coefficient $x=7.9$, absorbed Deuterium density $n_D
= 0.701\cdot 10^{22}\,$ cm$^{-3}$.
 Fig.~\ref{fig:Raiola2002Experiment}
 shows both $Y_{exp}^{\infty}$\/ and
the extracted $S_{exp}$, relative to the deuterated Tantalum
experiment of ref. \cite{Raiola:02}. The experimental results can be
reproduced only with an electron screening potential $U_e=309$\/ eV
$\gg U_{ad}=28$\/ eV;  similar large screening potentials are needed
to fit other deuterated-Tantalum experiments: $U_e=322$\/ eV in ref.
\cite{Czerski01} and $U_e=340$\/ eV in ref.
\cite{Raiola:02a,Bonomo:03}. The experimental procedure has been
critically studied in  \cite{Huke:07,Huke:08a}, especially target
surface contaminations and inhomogeneities in the implanted
deuterons distribution have been identified as possible sources of
systematic errors in the experimental data. However, performing
deuterated Tantalum targets experiments with different surface
contaminations, the electron screening potential needed to reproduce
the data is in the range $U_e=210-460$\/ eV \cite{Huke:07,Huke:08a},
much greater than $U_{ad}$\/ and in agreement with the previous
results. In the absence of incoming beam there are ions and
conduction electrons inside the metallic matrix, while the ion beam
may ionize deuteron atom target or metallic ions, so additional
charges may appear. Accurate numerical simulations performed by Huke
et al. \cite{Huke:08} suggest migration of electrons from the host
metal atoms to the Hydrogen ions during the impact event. This
effect produces a screening potential $U_e=39.7$\/ eV for
deuterated-Tantalum target, greater than $U_{ad}$\/ but still one
order of magnitude lower than the one needed to explain the
experimental results.

\sect{\label{sec:QuantumEff} Thermal effects and Quantum Uncertainty.}

\begin{figure}
  \begin{center}

        \includegraphics[scale=0.6]{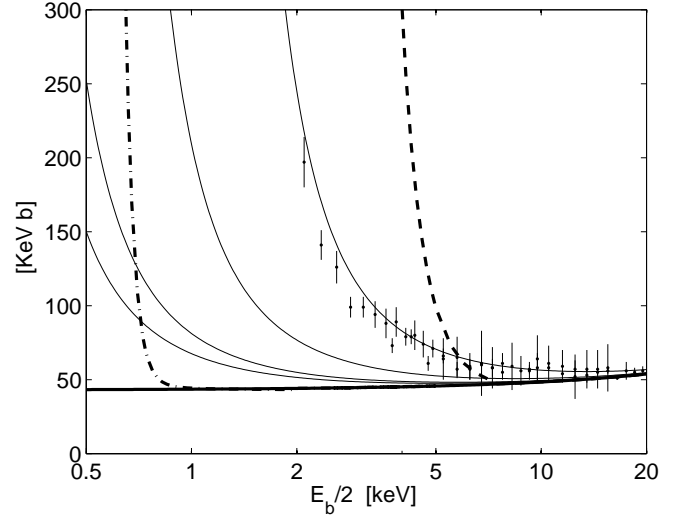}

  \end{center}
  \caption{\footnotesize Astrophysical Factor experimental points from ref. \cite{Raiola:02};
            $E_b$\/ is the incident Deuterons energy.
            Continuous curves show the zero-temperature Astrophysical factor $S$\/
            for screened (thin curves) and bare (thick curves) nuclei; here we adopted for $S$\/
            the same expressions of fig. \ref{fig:Raiola2002Experiment} with $E_{cm}=E_b/2$.
            The dashed and dot-dashed thick curves represent
            the thermal mean with Quantum Uncertainty Effect
            $S=\langle \sigma v_{rel} \rangle_{QE}\,
            \frac{m_D}{4}\sqrt{\frac{2E_b}{m_D}}\exp\left(\pi \sqrt{\frac{2E_G}{E_b}} \right)$\/
            in keV b, given by the numerical integration of
            eq.(\ref{eq:QEtherMeanInt}); specifically,
            the dot-dashed curve is relative to the Coulombian $\sigma_{coll}$\/, while the dashed one
            is plotted for the Ion Sphere Model $\sigma_{coll}$, eq. (\ref{eq:ISMcollCrossSec}),
            with $\alpha_1 =1$. }
   \label{fig:QErateRaiola2002Exp}
\end{figure}

Collision frequency among target and incoming deuterons is the most
important quantity to consider in this treatment. Collision
frequencies among other particles, e.g., target atoms and metallic
ions, are less important and are not responsible of modification of
momentum distribution function and of fusion rate \cite{Ferro:05}.
Momentum-energy dispersion relation is broadened by Quantum
Uncertainty Effect (QE), so that a Maxwell-Boltzmann energy
distribution $\Phi(E)\propto\exp(-E/k_bT)$ can yield corresponding
momentum distributions $\Phi(\mathbf{p}_t)$\/ with enhanced tails.
Since nuclear fusion cross section depends on the relative momentum
$\epsilon_{p_{rel}} = \frac{1}{2} \mu\, v_{rel}$\/ such an effect
increases the value of $\langle\sigma(\epsilon_{p_{rel}}) v_{rel}
\rangle /v$\/ respect to the one obtained with a sharp
energy-momentum relation. It is a good approximation of the
experimental situation to consider a beam of particles with definite 
energy and momentum and a Maxwell-Boltzmann energy distribution of 
thermalizedmtarget particles. The momentum-energy relation of the target
particle can be represented with a Lorentzian \cite{Ga:67}, at least
for $E_t\sim \epsilon_p$: \be
   \delta_\gamma(E_t - \epsilon_{p_t})= \frac{1}{\pi}\, \frac{\gamma}{(E_t - \epsilon_{p_t})^2 + \gamma^2}\; ,
   \label{eq:QEdispersionRel}
\ee where $\epsilon_{p_t} = \frac{1}{2} m_D\, v_t^2\,$, $\gamma =
\hbar\, n_D\, \sigma_{coll}\, v_{coll}\;$, $\; \sigma_{coll}$\/ and
$v_{coll}=\sqrt{2E_t/m_D}$\/ are the collisional cross section and
velocity. In
\cite{Starostin:2003next,Starostin:2003next2,Starostin:2005PhysUsp,Coraddu:06}
Quantum Uncertainty Effect has been applied to the fusion reaction
between beam and absorbed deuterons, adopting a Coulombian
collisional cross section $ \sigma_{coll} = e^4/\epsilon_{p_t}^2$.
This QE produces a rate increase, but only at beam energies lower
than the energy at which the experimental results start rising
($E_b\sim 2$\/ keV for ref. \cite{Raiola:02} instead of $E_b \sim
6-8$\/ keV, as shown in  fig. \ref{fig:QErateRaiola2002Exp}).

Zubarev \cite{Zubarev:06,Zubarev:07} proposes that the QE should be
effective  only for a small fraction of target deuterons that are in
a quasi-free mobile plasma states in the reaction zone and that
the rate enhancement is due to reactions of deuterons of this small
plasma fraction with the other stationary deuterons in the target.
The beam is only needed to maintain the plasma
states. Zubarev reproduces the enhancement with this mechanism and a
Coulombian collisional cross section.

However, collisional cross sections depend on the plasma
environment. In particular,  the Coulombian cross section $
\sigma_{coll} = e^4/\epsilon_{p_t}^2$\/ is appropriate only for
weakly interacting plasmas, i.e., plasmas with parameter $\Gamma =
\frac{e^2}{k_bT a_{ws}}\ll 1$, where $a_{ws} = (3/4\pi n_D)^{1/3}$\/
is the Wigner-Seitz radius, but $\Gamma\sim 100$ for the typical
experimental conditions we are interested in,  see for instance ref.
\cite{Raiola:02}, where $n_D=0.701\cdot 10^{22}$\/ cm$^{-3}$,
$k_bT=0.0244$\/ eV, $a_{ws}=3.24\cdot 10^{-8}$\/ cm, resulting in
$\Gamma = 182$\/ (very close to the liquid-solid transition).
Therefore, a strong coupled plasma scheme should be
used for the absorbed deuterons. A model that can describe the
screened collisional cross section in this limit is the Ion Sphere
Model \cite{Ichimaru:1992}:
 \be
    \sigma_{coll}\, =\, 2\pi\, \alpha_1^2\, a_{ws}^2
    \label{eq:ISMcollCrossSec}\; .
\ee Eq. \ref{eq:ISMcollCrossSec} is a non trivial expression for
$\sigma_{coll}$, because the parameter $\alpha_1$\/ contains
information on the particle-particle correlations and in ref.s
\cite{Ichimaru:1982,Ichimaru:1993} (and citation therein) Ichimaru
and collaborators have given a detailed analysis of its expression
and evaluation. For what concerns the present problem we can say
that $\alpha_1$\/ is a correlation factor of the order of unity or,
more precisely, that its value lies between 0.4 and 0.9.

The resulting  distribution for $\mathbf{p}_t$ is
$\Phi(\mathbf{p}_t) = \frac{1}{I_N}\, \int_0^\infty d E_t\,
\delta_\gamma(E_t - \epsilon_{p_t}) e^{-E_t/k_bT} $, where $I_N =
4\pi \int_0^\infty p_t^2 dp_t \int_0^\infty d E_t\,
\delta_\gamma(E_t - \epsilon_{p_t}) e^{-E_t/k_bT}$\/ is the
normalization integral. Therefore, the thermal mean is:
\bear
  \lefteqn{ \langle \sigma v_{rel} \rangle_{QE}\,  = \,  \int d^3\mathbf{p}_t \Phi(\mathbf{p}_t)\, \sigma v_{rel}\, 
                                                =\,\frac{2\pi m_D^{2/3}}{I_N} \int_{-1}^{+1} d\cos\vartheta }  \nonumber  \\
                                       &  & =\,  \int_0^\infty d\epsilon_{p_t}\, \sqrt{2 \epsilon_{p_t}}
                                                \int_0^\infty d E_t\, \delta_\gamma(E_t - \epsilon_{p_t})
                                                 e^{-E_t/k_bT} \sigma (\epsilon_{p_{rel}}) v_{rel}\; , \nonumber \\
                                      \label{eq:QEtherMeanInt}
\eear 

where $\vartheta$\/ is the angle between $\mathbf{p}_t$\/ and
the beam and $v_{rel} = \sqrt{2/m_D}\, (E_b + \epsilon_{p_t} -2\sqrt{E
\epsilon_{p_t}}\cos\vartheta)^{1/2}$.

 Eq. (\ref{eq:QEtherMeanInt}) has been numerically evaluated with
 parameters appropriate to the experiment of ref. \cite{Raiola:02}
 using a constant
Astrophysical factor $S(\epsilon_{p_{rel}})\simeq S_0 = 43$\/ keV b,
valid within $\approx 6\%$\/ for $E_b\leq$\/ 10 keV. The
$\vartheta$\/ integration has been done analytically in terms of
incomplete Gamma-Euler function, while the remaining integrations
have been performed using the Gauss adaptive method; results are
shown in fig. \ref{fig:QErateRaiola2002Exp} for Coulombian and Ion
Sphere Model (ISM) collisional cross sections. Both cases yield a
strong enhancement of the reaction rates at low-energy: below
$E_b\sim 2$\/ keV for the Coulombian and below $E_b\sim 10$\/ keV
for the ISM case. These behaviors should be compared to the
anomalous enhancement of the experimental data that start below
energies $E_b\sim 4-6$\/ keV. In the ISM case, the energy-threshold
below which the rate is enhanced depends on $\sigma_{coll}$  that,
in turns, depends on $n_D^{-2/3}$, (see eq.
(\ref{eq:ISMcollCrossSec})).
 Our calculation uses no adjustable parameter ($\alpha_1=1$\/ in fig.
\ref{fig:QErateRaiola2002Exp}), however, if we follow the approach
of Kim and Zubarev in ref. \cite{Zubarev:07}, we could apply the
enhancement only to that fraction of the absorbed deuterons that is
in a quasi-free plasma state.
 The dependence of the reaction rate $r$ on the fraction
$f=n_{D_m}/n_D$ of absorbed target Deuterons, $n_D$, that are
quasi-free, $n_{D_m}$, is not trivial: $r$ is not simply
proportional to $f$, but it has an additional dependence on $f$
through the ISM collisional cross section $\sigma_{coll}$\/  eq.
(\ref{eq:ISMcollCrossSec}); it is clear, however, that $f=1$ gives
the maximal effect. In principle, the fraction $f$ could be
calculated in some theoretical model, determined from an independent
experimental measurement, or used as a free parameter to be fitted.
 Results presented have been obtained with $f=1$ and, therefore, should be
 interpreted as upper limits for this effect.

\sect{\label{sec:DefDistr} Modified Debye-H\"{u}ckel screening.}

\begin{figure}
 \begin{center}

        \includegraphics[scale=0.6]{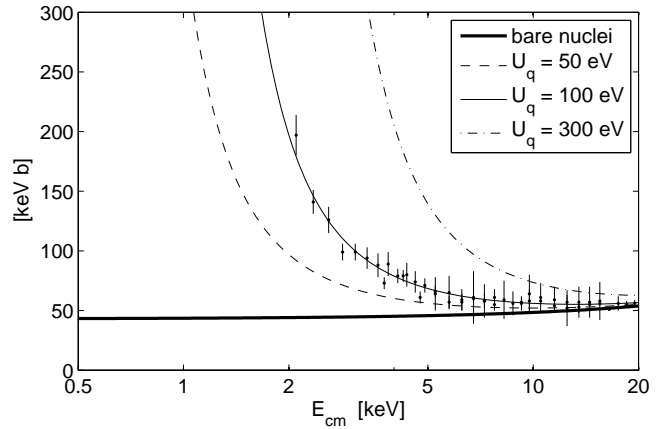}

 \end{center}
  \caption{\footnotesize   Astrophysical Factor experimental points from ref. \cite{Raiola:02}.
            Bare nuclei curve correspond to $S_{bare} = 43 + 0.54 E_{cm}$\/ keV b,
            while the screened curves are $S=f_q\cdot S_{bare}( E_{cm})\,$ ,
            where $f_q$\/ is defined in eq.(\ref{eq:ModifDHpotCorrFact}),
            $U_q = e^2 \langle \frac{1}{R_{DH}} \rangle\, $ , $g(q)=3-2q$\/ and $q=0$.
            Ion thermal motion is neglected: $E_{cm}=E_b/2$. }
  \label{fig:ModifDHRaiola2002Exp}
\end{figure}

 We discuss the effect of electron screening on deuteron-fusion
reactions in metal matrix. We do not consider the interplay with the
effect discussed in the previous section.
The screening potential
at distances near and below the turning point influence the most
reaction rates.
Thermal effects and screening phenomena in plasma environment are
strictly connected. Assuming, in a ideal plasma, a Maxwell-Boltzmann
(M-B) thermal distribution, the Debye-H\"{u}ckel ion-screening
potential \( V_{DH}(r) = \frac{Z_1Z_2 e^2}{r} \exp (-r/R_{DH}) \)\/
is obtained by means of the linear Poisson equation ($R_{DH} =
\sqrt{k_bT/4\pi n Z e^2} $\/ is the Debye-H\"{u}cke radius). In
strongly coupled plasmas, Debye description loses its physical
interpretation. In the case studied, the main reason  is that the
number of particles inside the Debye sphere is too small. The D-H
approach needs to be extended to include features that appears as
systems leave the weak-coupled regime. Near the weak-coupled regime
deviations from the D-H regime can be parameterized by an electron
cloud with a steady-state generalized spatial distribution of q-type
(for $q\rightarrow 1$\/ electrons are distributed according to a
Boltzmann factor): we assume that such deviation can be analytically
continued in the strong-coupled regime.
 Following this strategy, Quarati and Scarfone
\cite{Quarati:07} have recently derived a new screening potential
called Modified Debye-H\"{u}ckel potential using two different
approaches. The first one uses a generalized non-linear Poisson or
Bernulli equation, the second is based on superstatistics
\cite{Beck:01,Beck:04}. We discuss this second approach in some
detail in the following.\\
We choose the value $q=0$, also for the Ta matrix, because we can
reproduce an electron distribution spatially concentrated around the
deuteron, with a strongly depleted tail with cut-off at $R_{DH}/3$.
We are encouraged in this
line of treatment also by the result of ref. \cite{Saltz:08}.\\
The authors of ref. \cite{Quarati:07} assumed that non-linear effects produce
fluctuations on the inverse
Debye-H\"{u}cke radius $1/R_{DH}$, with a Gamma-function probability distribution:
\(
   f_{q} (r, \lambda, \lambda_0)\, =\, \frac{A_q(r, \lambda_0)^{\frac{1}{1-q}}}{\Gamma \left(\frac{1}{1-q} \right)}
                                       \lambda^{\frac{1}{1-q} - 1} e^{-\lambda A_q(r, \lambda_0)}    \;
\),
where $f_{q} (r, \lambda, \lambda_0)$\/ represents the probability density to
observe a certain value $\lambda$\/ spreads around a central value
$\lambda_0$.
To obtain from $f_{q} (r, \lambda, \lambda_0)$\/
an electron depleted tail distribution with a cutoff,
we limit the entropic index $q$\/ into
the $0\leq q \leq 1$\/ interval,  assuming:
\[
   A_q(r, \lambda_0)\, =\, \frac{1}{(1-q)\, g(q)\, \lambda_0 }\, -\, r\; ,
\]
where $ g(q)$\/ is a generic entropic index function, that satisfies
the condition $ g(1)=1$\/ (in ref. \cite{Quarati:07} the choice  $
g(q)=(2-q)^{-1}$\/ is adopted).
The point charge potential $V_q(r)$\/ can be identified by the functional:\\
\(   \mathcal{F}_q (r, \lambda_0)\, =\,  C_q \int_0^\infty  f_{q} (r, \lambda, \lambda_0) e^{-\lambda r}\, d\lambda \; \),\\
through the relation:
\be
   r\, V_q(r)\, =\,  \frac{D}{C_q \left\langle\frac{1}{R_{DH}}\right\rangle}\, \mathcal{F}_q (r, \lambda_0)\; ,
   \label{eq:PotentialFunctionalRel}
\ee
where $C_q= (2-q) g(q)$\/ is a normalization factor,
$D=Z_1 Z_2 e^2 \left\langle \frac{1}{R_{DH}} \right\rangle$\/
and $\lambda_0 =\langle\lambda\rangle
               =  \int_0^\infty f_{q} (r, \lambda, \lambda_0)  \lambda\, d\lambda
               = \left\langle \frac{1}{R_{DH}} \right\rangle$.
The charged particles distribution $\rho(r)$\/ and the point charge
potential $V_q(r)\propto \rho(r)$\/ can be derived from eq.
(\ref{eq:PotentialFunctionalRel}), developing the functional
$\mathcal{F}_q (r, \lambda_0)$\/ with the previous assumption. A
Tsallis cut-off form \cite{Tsallis:88,Tsallis:03} is obtained for
the potential: \be
  V_q(r)\, =\, \left\{ \begin{array}{l}
                       \frac{Z_1 Z_2 e^2}{r}
                       \left( 1\, -\, (1-q)g(q) \left\langle \frac{1}{R_{DH}} \right\rangle\, r \right)^{\frac{1}{1-q}}\; , \\ 
                       \begin{array}{lcr}
                                         &  \;\;\;\;\;\;\;\;\; \mbox{if}  &
                        r < \frac{1}{(1-q)g(q) \left\langle \frac{1}{R_{DH}} \right\rangle},  \\
                                                                                              \\
                                     0   &  \;\;\;\;\;\;\;\;\; \mbox{if}  &
                        r \geq \frac{1}{(1-q)g(q) \left\langle \frac{1}{R_{DH}} \right\rangle}\; .
                        \end{array}
                        \end{array}
                \right.
  \label{eq:DebHuckModifPotRevised}
\ee 
We note the relation of the power-law expression of $V_q(r)$\/
with Tsallis-like distributions. In systems with long-range
interactions and correlations and/or memory or in systems that are
not in global thermodynamic equilibrium, but rather in a metastable
state with a long, finite lifetime, such deformed Tsallis-like
distributions are relevant at the very least as convenient
parameterizations of deviations from equilibrium; superstatistics is
an example of such an  approach~\cite{Beck:01,Beck:04}.\\
 The main contribution to the charged
particles fusion cross section is given by the screening barrier
penetration factor $P(E)$. In the standard Debye-H\"{u}ckel
potential case, the simplified expression $\sigma(E) =
\frac{S(E)}{E}P(E)$\/ can be obtained, that differs from the bare
nuclei cross section of eq. (\ref{eq:CrosSecbareNuc}) only for the
penetration factor $P(E)=\exp\left( -\pi \sqrt{\frac{E_G}{E+U_{DH}}}
\right)$, where $U_{DH}=\frac{Z_1 Z_2 e^2}{R_{DH}}$. In the case of
the modified Debye-H\"{u}ckel potential $V_q(r)$, the penetration
factor $P_q(E)$\/ is given by the expression: \be
  P_q(E)\, =\, \exp\left[ -\frac{2}{\hbar c}
                          \int_0^{r_q} \left[ 2 \mu c^2 (V_q(r) - E)\right]^{\frac{1}{2}} dr \right]\; ,
  \label{eq:PenBarrierDHmodifScreen}
\ee
where the classical turning point $r_q$\/ has to be determined through the equation
$V_q(r_q) = E$.\\
Eq. (\ref{eq:PenBarrierDHmodifScreen}) can be analytically solved in
the  $q=0$\/ case, obtaining $r_q = Z_1 Z_2 e^2 / (E+g(0)D)$\/ and
$P_0(E) = \exp\left( -\pi \sqrt{\frac{E_G}{E+g(0) D}}  \right)$.\\
The bare nuclei cross section reported in eq. (\ref{eq:CrosSecbareNuc})
can be corrected, to account the modified Debye-H\"{u}ckel (D-H) screening,
multiplying $\sigma_{bare}(E)$\/ by the factor
\be
    f_q\, =\,   P_q(E)\, \exp \left( \pi \sqrt{\frac{E_G}{E}}  \right)
    \label{eq:ModifDHpotCorrFact}
\ee For instance we compare the Modified D-H Astrophysical Factor
for D+D reaction with the  ref. \cite{Raiola:02} experimental data,
adopting the choice $g(q)=3-2q$. The results is shown in fig.
\ref{fig:ModifDHRaiola2002Exp} for the entropic index $q=0$. One can
observe as a screening potential $U_q=D$, three times lower than in
the standard D-H case ($U_q\sim$ 100 eV instead of $U_{DH}\sim$ 300
eV), is required to reproduce the experimental data. An
electrostatic screening potential of this order of magnitude has
been obtained, for instance, by Saltzmann and  Hass \cite{Saltz:08}
through a Thomas-Fermi model of the electron gas in a
deuterated-copper target (they obtained a screening potential of 163
eV, instead of
the 470 eV needed to reproduce the experimental results).\\
Now we are treating $g(q)$\/ and $q$\/ as free parameters, but, in
principle, a link can be establish between the inverse D-H Radius,
the temperature fluctuations and the $q$-index: \(
\frac{\Delta\left( \frac{1}{R_{DH}} \right)}{ \frac{1}{R_{DH}}}
                 \, =\, \frac{\Delta (k_bT)}{k_b T} \, =\, \sqrt{1-q} \;\)
and by the electron charge $q$-distribution around the ion
(if experimentally known or as deduced from a model).
By this way the modified  Debye-H\"{u}ckel potential can be obtained
starting from the environment condition.

\sect{\label{sec:Conc}  Conclusion}

We studied two effects of non-Maxwellian velocity distributions in
plasmas as possible explanations for the fusion rate enhancements
observed in deuterated-metal target experiments: modifications to
the thermal mean $\langle \sigma v_{rel} \rangle$ due to the Quantum
Uncertainty Effect and the stronger screening arising from
non-thermal electron distributions. Either effect reduces the
discrepancy between theoretical models and experimental results.

The broadening of the momentum distribution due to the QE has been
studies using a collisional cross section $\sigma_{coll}$\/ derived
by the Ion Sphere Model, eq. (\ref{eq:ISMcollCrossSec}), model that
captures the main features of collisions in the strong coupled
plasmas  that characterize the experimental environment.

We have been able to obtain a rate enhancement up to energy
thresholds much higher than previous calculations \cite{Coraddu:06}
and even above the one observed in experiments, as shown in fig.
\ref{fig:QErateRaiola2002Exp}. If one makes the reasonable
additional hypothesis that  only a fraction of the absorbed
deuterons are in a quasi-free plasma-state, which can be shown to be
still strongly coupled, one should be able to reproduce the
experimental behavior.  We intend to explore more quantitatively
this possibility in the near future.

We have also considered the effects of plasma screening on the
reaction cross sections, when the modified Debye-H\"{u}ckel
potential introduced by Quarati and Scarfone in ref.
\cite{Quarati:07} is used. Already in ref. \cite{Saltz:08}  a
screening potential $U_e\sim 100$\/ eV was obtained, greater than
the adiabatic limit $U_{ad}$\/, but still too low to reproduce the
experimental results. In this work we showed that the modified D-H
potential of eq. (\ref{eq:DebHuckModifPotRevised}), with an
appropriate choice of the function $g(q)$\/ and of the entropic
index $q$, can reproduce the data, see in fig.
\ref{fig:ModifDHRaiola2002Exp}. In principle the value of the
entropic index $q$\/ can be derived from the plasma equation of
state and $g(q)$\/ from the electron charge distribution around the
ion, then the modified D-H screening contribution to the rate
enhancement can be evaluated without free parameters. We shall
investigate this last point in a future work.

\end{document}